\documentclass{elsart}
\usepackage{amsmath}
\usepackage{epsfig}
\usepackage{amssymb}

\begin{document}

\begin{frontmatter}



\title{On the behavior of micro-spheres in a hydrogen pellet target}

%
%
\author[ISV]{{\"O}.~Nordhage\corauthref{cor}},
\ead{orjan.nordhage@tsl.uu.se}
\author[IMPCAS]{Z.-K.~Li\corauthref{cor}},
\ead{lizhankui@impcas.ac.cn}
\author[TSL]{C.-J.~Frid{\'e}n},
\author[TSL]{G.~Norman}, and
\author[ISV]{U.~Wiedner}

\corauth[cor]{Corresponding authors.}
\address[ISV]{Department of Radiation Sciences, Uppsala University, Box 535, SE-751 21 Uppsala, Sweden}
\address[IMPCAS]{Institute of Modern Physics, the Chinese Academy of Science, Lanzhou, 730000, China}
\address[TSL]{The Svedberg Laboratory, Uppsala University, Box 533, SE-751 21 Uppsala, Sweden}

\begin{abstract}
A pellet target produces micro-spheres of different materials,
which are used as an internal target for nuclear and particle
physics studies. We will describe the pellet hydrogen behavior by
means of fluid dynamics and thermodynamics. In particular one aim
is to theoretically understand the cooling effect in order to find
an effective method to optimize the working conditions of a pellet
target. During the droplet formation the evaporative cooling is
best described by a multi-droplet diffusion-controlled model,
while in vacuum, the evaporation follows the (revised)
Hertz--Knudsen formula. Experimental observations compared with
calculations clearly indicated the presence of supercooling, the
effect of which is discussed as well.
\end{abstract}

\end{frontmatter}


\section{Introduction}
The usage of internal targets in storage-ring accelerators has
opened up a new era for the investigations of nucleon-nucleon
collisions with high precision in hadron physics. The figure of
merit for physics experiments is usually the (integrated)
luminosity, i.e.\ the target thickness times the particle beam
intensity. For light targets such as hydrogen or deuterium it is
quite difficult to achieve the necessary high luminosity with a
conventional gas target. The alternative of using frozen
micro-spheres, so-called pellets, has been suggested and a
prototype pellet-generator was developed by Trostell \cite{Tr95}.
A pellet target avoids unnecessary background gas load to the
accelerator ring and allows an effective target thickness of a few
times $10^{15}\,$atoms/cm$^{2}$ \cite{Ek95,Ek02}. Since the target
tubes connected to the accelerator ring are rather narrow,
detectors can be placed close to the interaction point in a nearly
$4\pi\,$sr configuration. These particular characteristics of a
pellet target have caused worldwide attention, and in the near
future pellet target facilities are foreseen in J{\"{u}}lich,
Germany, and in Lanzhou, China. Furthermore, this kind of target
is considered as one of the main options for the future PANDA
experiment \cite{LoI} at FAIR in Darmstadt, Germany.

Until now there is only one pellet target facility permanently
installed and actively operated inside a storage-ring accelerator.
It is connected to the WASA (Wide Angle Shower Apparatus) detector
at CELSIUS at The Svedberg Laboratory, Uppsala, Sweden. This
target has successfully provided hydrogen pellets for experimental
data taking since 1999, with deuterium pellets being produced as
well since 2004.

The principle of operating a pellet target has been described
elsewhere \cite{Tr95,Ek96}, and here we only briefly review those
parts which are of importance for this work. Compared to the
earlier work, we have undertaken more systematic studies to
theoretically understand and optimize the target performance. We
will describe the behavior of a pellet target on the basis of a
thermodynamic analysis. We will further restrict ourselves to
hydrogen pellets, but the mechanism is similar for deuterium
pellets. Data values concerning hydrogen are taken from
Ref.~\cite{Souers86}, as throughout this work. Concerning the
micro-spheres, the naming convention used in this paper is
\emph{droplet} whenever we talk about the liquid phase and
\emph{pellet} for the solid phase, while the word
\emph{micro-sphere} covers both cases as well as the intermediate
one.

\section{Experimental setup}
The pellet-generation system consists of four parts:
\begin{enumerate}
\item a coldhead with heat exchangers, in which pressurized
hydrogen gas is cooled and liquified, \item a droplet formation
chamber (DFC), where a glass nozzle together with an acoustical
excitation system produces and breaks-up the jet of liquid
hydrogen into uniformly spaced and sized droplets, \item a vacuum
injection system through which the droplets transform to totally
frozen micro-spheres (and hereafter are called pellets), and \item
a skimmer to collimate the pellet beam.
\end{enumerate}

\begin{figure}[htb]
\begin{center}
\includegraphics*[height=7cm]{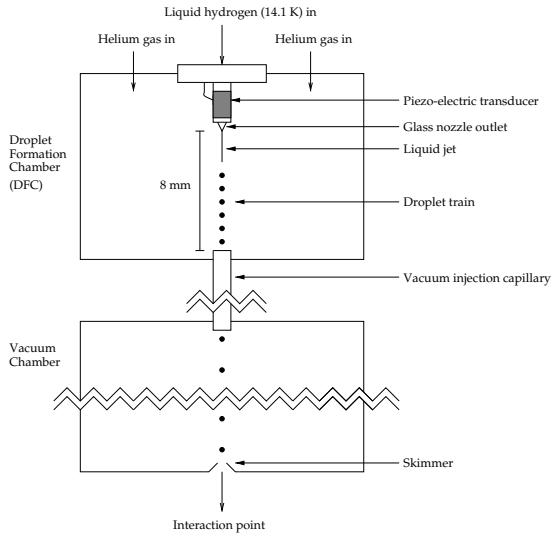}
\end{center}
\caption{A schematic picture of the droplet formation chamber and
the vacuum chamber. The geometry of the DFC is complex but can be
approximated by a cylinder of diameter $22\,$mm and height
$15\,$mm. The distance between the nozzle outlet and the inlet of
the vacuum injection capillary is $8\,$mm. The distance between
the outlet of the vacuum injection capillary and the skimmer is
$70\,$cm.} \label{fig:DropFormChamber}
\end{figure}

The hydrogen is cooled by a commercially available two-stage
coldhead. At stage 1, a first heat exchanger is mounted to
cool the gas to about $50\,$K. At stage 2, a second heat
exchanger regulates the temperature to $14.1\,$K. Inside this
latter heat exchanger the gas will be liquified and brought to the
nozzle directly attached at the exit, as depicted in
Fig.~\ref{fig:DropFormChamber}. The carrier gas, helium, which is
brought to the DFC in a separate channel is also cooled by this
heat exchanger though it will always stay in gas phase.

\begin{figure}[htb]
\begin{center}
\includegraphics*[height=7cm]{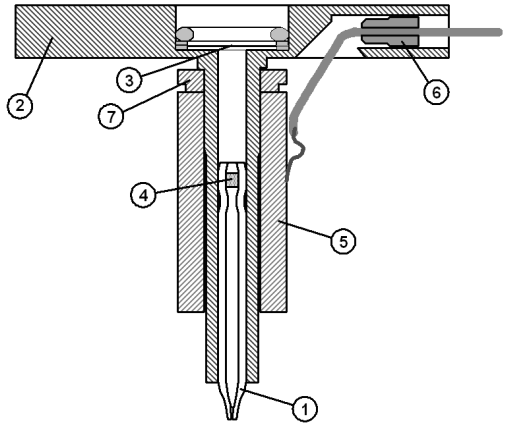}
\end{center}
\caption{A sectional drawing of the nozzle unit; showing 1) the
glass nozzle, 2) the copper holder, 3) the Millipore filter
($0.5\,\mu$m), 4) the stainless steel filter ($2.0\,\mu$m), 5) the
piezo-electric transducer, 6) the electric feed through to this,
and 7) the kovar washer.} \label{fig:LJN}
\end{figure}

The breakup of the liquid hydrogen jet is a result of the
acoustical excitation with axial symmetry induced by the ceramic
piezo-electrical transducer connected to the nozzle (see
Fig.~\ref{fig:LJN}). The DFC is a stainless steel chamber with
four windows allowing optical observations in both transverse
directions, which enables us to control the droplet formation.

The vacuum injection consists of a $70\,$mm long capillary,
connected to a differentially pumped vacuum chamber. Another
$70\,$cm downstream the exit of the vacuum injection capillary, a
collimating skimmer is placed to skim off the pellets with too
large angular divergence. The collimated pellets will proceed an
additional $1.41\,$m before they reach the interaction point with
the ion beam. As a last step they are captured by active charcoal
inside the cryogenic beam dump. The mechanical details of this
facility can be found in Ref.~\cite{Tr95}.

\section{Operating conditions and measurements}
The operating conditions of the pellet target have been developed
steadily over the years. The inner diameter of the nozzle exit has
decreased from initially roughly $17\,\mu$m to presently
$12\,\mu$m, which in turn has resulted in a pellet diameter
decrease from $50\,\mu$m to a value of about $30\,\mu$m. The rate
of hydrogen pellets reaching the interaction point has increased
from $3 \times 10^{3}\,$s$^{-1}$ to almost $10^{4}\,$s$^{-1}$. The
optimal working frequency of the transducer is conditionally
determined and may vary over a wide range, typically from
$60\,$kHz to $100\,$kHz. Nevertheless, for an individual
experimental run the working frequency can only change within a
quite narrow frequency window. It can only be determined
experimentally by the observations of a ``nice'' droplet train, a
good micro-sphere survival ratio through the vacuum injection
capillary, and a strong concentration of pellets at the center of
the skimmer. In this article, the logged data from the last
hydrogen run in December 2003 were used as example. The conditions
for this run were: a measured distance of $0.368\,$mm between two
successive droplets in the droplet formation chamber, a transducer
frequency of $102.32\,$kHz, a nozzle diameter $12.0\,\mu$m, a
pressure in the droplet formation chamber of $21.3\,$mbar, and a
driving pressure of $729\,$mbar for the injected hydrogen gas.

Normally, either cooled hydrogen gas or helium gas at a
temperature of about $17\,$K could be used as the background gas
inside the DFC. To get a good pellet concentration at the skimmer
one wants to have as low pressure as possible in the DFC,
otherwise the angular pellet divergence from the vacuum injection
becomes too large. If the absolute pressure in the droplet chamber
is too low there is a risk that the liquid jet will degenerate
into a spray which immediately freezes and blocks the jet. This
risk has been significantly reduced by the use of helium as
background gas. The higher viscosity of helium gives a more steady
gas flow through the vacuum injection capillary. As a result of
this it is possible to have a higher absolute pressure in the DFC
and still get a good concentration at the skimmer. The lower
critical temperature of helium also eliminates the risk that the
gas will liquefy in the second heat exchanger.

The ambient gas in the DFC is thus a mixture of two kinds of
gases; helium and the evaporated hydrogen gas. The flow of helium
is regulated by the total DFC pressure. In addition, it was found
that when the helium supply was shut off, the background gas which
only consisted of hydrogen had an equilibrium pressure of
$8\,$mbar. Thus we assume that in the gas mixture the hydrogen
vapor was also occupying a partial pressure of $8\,$mbar, and the
residue $13\,$mbar is ascribed to helium gas for a constant total
pressure of $\sim21\,$mbar in the DFC.

It has been theoretically and experimentally proven that the
hydrogen droplets are fragile because of their rather low surface
tension and thus are easily destroyed by the high-speed gas flow
in the vacuum injection capillary while solid hydrogen-spheres
would probably survive the vacuum injection better
\cite{HA63,HA72}. To avoid freezing of the liquid jet itself but at the
same time enable the droplets to freeze as soon as possible, the
temperature of liquid hydrogen at the exit of the nozzle is
precisely controlled to be a slightly higher than the normal
freezing temperature, $T_\mathrm{m}=13.96\,$K, namely $14.1\,$K,
while the total pressure, thus also the partial pressure of
hydrogen vapor, in the droplet formation chamber is set to a
value much lower than the triple-point pressure of hydrogen
($P_\mathrm{tp}=72\,$mbar).

\begin{figure}[htb]
\begin{center}
\includegraphics*[height=7cm]{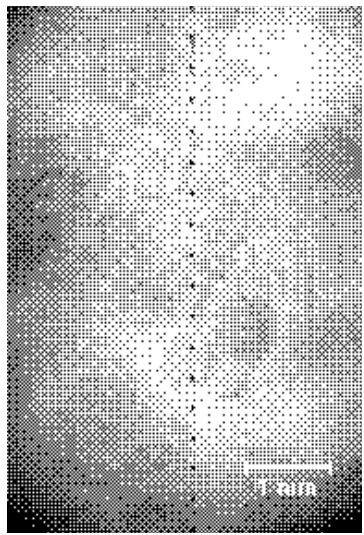}
\end{center}
\caption{A close-up of the droplet train in the droplet formation
chamber showing both satellites and main droplets.}
\label{fig:DropletTrain}
\end{figure}

Pictures (see Fig.~\ref{fig:DropletTrain}) of the droplets have
been taken in the experiments by a digital camera with the
assistance of a stroboscopic flash diode. The measured
inter-droplet distance $\lambda$ was 0.368 mm which consequently
resulted in a droplet diameter of $D_\mathrm{d}=39.1\,\mu$m from
Eq.~(\ref {eq:Dd_Djet}). This is also consistent with the measured
value of the ratio $\lambda /D_\mathrm{d}\simeq 9$, when the
droplets are just detached from the jet.

The droplet velocity $v_\mathrm{d}$ is deduced from
$v_\mathrm{d}=\lambda f$, where $f$ is the driving frequency of
the transducer. For our case we find $v_\mathrm{d}=37.7\,$m/s. The
observed relative velocity decrease for droplets is measured to be
about $6\%$, comparing droplets just detached from the jet to
those close to the inlet of the vacuum injection capillary.

\begin{figure}[htb]
\begin{center}
\includegraphics*[height=7cm]{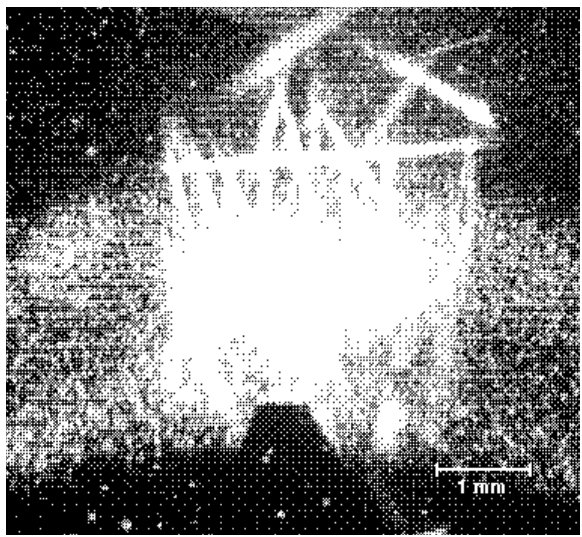}
\end{center}
\caption{Shown is the pellet concentration lit by a laser just
above the skimmer, visible as the black cone in the lower part.
Its opening diameter is $0.59\,$mm. We also see the traces of
pellets bouncing around, confirming that they really are solid at
this location, i.e.\ $70\,$cm downstream of the outlet of the
vacuum injection capillary. The horizontal pattern comes from the
monitor which the photo is taken of. This also makes the pellet
concentration to look significantly bigger compared to the
original frames from the CCD camera.} \label{fig:skimmer}
\end{figure}

From experimental observations we know that the micro-spheres are
in liquid phase at the inlet of the vacuum injection capillary and
completely frozen at the skimmer. The latter conclusion comes from
observing the micro-spheres that do not pass to really rebounce
like billiard balls, i.e.\ solid objects, as indicated in
Fig.~\ref{fig:skimmer}. The first conclusion arises from tilting
the droplet train to hit the capillary inlet. In this case we do
not observe the ``egg shells'' as described in
Ref.~\cite{Foster77} which should have been the result if the
micro-spheres been partially frozen. To conclude our observations,
the micro-spheres must freeze somewhere in between the inlet of
the vacuum injection capillary and the skimmer.

The calculations (see Fig.~\ref{fig:T_vs_t}) show that the
hydrogen droplets are cooled down so fast by evaporative cooling
that a solid micro-sphere (pellet) or at least a solid shell at
the droplet surface could be formed at the exit of the droplet
formation chamber (inlet of vacuum injection capillary). Clearly a
phenomenon called supercooling is inevitably encountered, as has
been mentioned before but not examined in greater detail
\cite{Tr95}. It is still not clear how much of the pellet beam
divergence is ascribed to the supercooling, but as mentioned the
pellet survival rate may be deeply affected by this phenomenon.

The gas flow in the vacuum injection capillary will be ``choked"
due to the large pressure difference in the DFC and the vacuum
chamber. The droplet carried by the gas flow will endure a drag
force given by Eq.~(\ref{eq_f_drag}) and as a result the droplet
will be accelerated. We have experimentally measured the average
pellet velocity of hydrogen from the pictures of the pellets just
after the vacuum injection capillary. For this particular run, the
average over 52 distances together with the transducer frequency
resulted in a pellet velocity of $94\,$m/s. For other runs, the
typical values range from $60\,$m/s to $100\,$m/s dependent on the
pressure in the DFC and the driving pressure of hydrogen gas to
the nozzle. At the moment the distribution of the pellet velocity
is unknown, but we will report on these experiments in a
forthcoming paper.

A crucial topic originating from the vacuum injection is the
pellet beam divergence which causes about 80 percent of the
produced hydrogen pellets to be skimmed off because of their too
large angular spread, i.e.\ most of them never reach the
interaction point. The origin of this type of divergence is not
known in detail \cite{Meyer90}.

\begin{figure}[tbh]
\begin{center}
\includegraphics*[height=7cm]{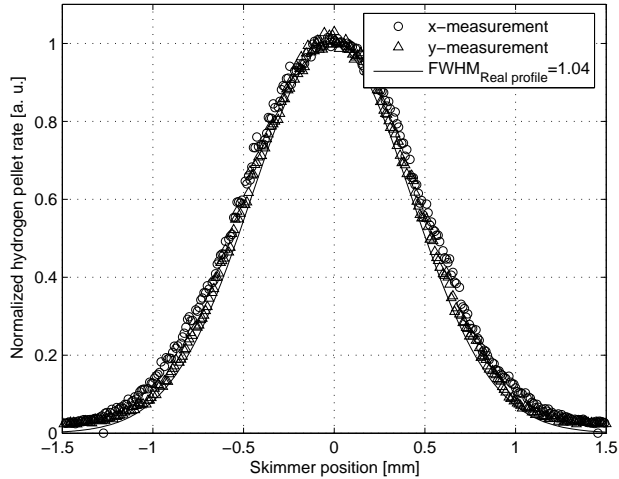}
\end{center}
\caption{Measured hydrogen pellet beam profile at the skimmer,
i.e.\ $70\,$cm from the vacuum injection capillary exit. The scans
are carried out in both directions. The skimmer diameter was
$0.59\,$mm and the corrected symmetric Gaussian for the real beam
had a FWHM of $1.04\,$mm.} \label{fig:BeamProf_H2}
\end{figure}

Pellet beam profiles measurements were carried out by step-wise
moving the pellet beam over the skimmer aperture while counting
the ones that pass. The results are shown in
Fig.~\ref{fig:BeamProf_H2}. \emph{A priori} the shape of the
uncollimated beam, i.e.\ before the skimmer, is unknown. However,
by fitting several test-functions to the experimental results a
Gaussian shape reproduces the measured values the best. Since the
size of the pellet beam is similar to the skimmer diameter, the
convolution between the two becomes important. We assumed the
uncollimated beam to be a symmetric Gaussian, and by fitting its
convoluted result to the experimental measurements the ``real''
pellet beam profile, with a FWHM of $1.04\,$mm, was obtained.

From comparing the total number of pellets in the uncollimated
beam to the transducer frequency, a survival ratio of 50\% was
deduced. This rather low number is also supported from the photos
taken just below the vacuum injection capillary. They indeed show
that the inter-pellet distance is sometimes twice the distance of
others. Thus a pellet in such a case is ``missing''.

\section{Interpretation of the results}
\subsection{Droplet formation} It is well known that pressurized liquid
ejected from a nozzle will form a jet. A cylinder of jet is
dynamically unstable under the action of surface tension. When a
vibration of a certain frequency is applied to the nozzle the
corresponding wavelength $\lambda$ will force the jet to
disintegrate into a stream of uniform-sized droplets. From Lord
Rayleigh's analysis the disturbance to the shape of the droplets
grows most rapidly at an approximate wavelength of $\lambda=4.5
D_\mathrm{jet}$, where $D_\mathrm{jet}$ is the diameter of the jet
\cite{Ray79,Schneider64}. In our case the working frequency has to
be set at a much lower frequency than this optimal one to avoid
the coalescence effect causing clustering in the high-frequency
mode. We have observed experimentally, like others \cite{PiLe77}
also have, that satellite droplets are inevitably formed
accompanying the main droplets, quite independent of the
experimental configuration. However, we can tune the transducer
frequency slightly so that the satellite droplets could merge into
the main droplets as quickly as possible due to their different
velocities. For details of the nozzle unit, see
Fig.~\ref{fig:LJN}.

\subsubsection{Droplet size determination}
\label{sec:D_size} To determine the droplet size one needs to know
the jet diameter. However, the small jet size for a 12-$\mu$m
nozzle in a cryogenic environment is difficult to be measured
directly. We apply fluid dynamics to overcome this problem.

In fluid dynamics the Reynolds number, defined as
$\mathrm{Re}=\rho v_\mathrm{noz}D_\mathrm{noz}/\mu$, is used to
distinguish between different types of flows. Here, $\rho$ is the
density of liquid hydrogen, $v$ is the fluid velocity, and $\mu$
is the dynamic viscosity. When $\mathrm{Re}\lesssim 2000$ the flow
is always laminar, and when $\mathrm{Re}>4000$ it is always
turbulent \cite{Meyer90}. For a $12.0\,\mu$m inner-diameter nozzle
we get $\mathrm{Re}\simeq 36v$ for hydrogen, which means that in
order to have a laminar flow the hydrogen fluid must have a
velocity $\lesssim55\,$m/s. This is indeed the case, because of
the discussion in Sec.~\ref{sec:D_vel_ev}, resulting in a Reynolds
number of about 1000.

Most of the pressure drop between the driving pressure and the
much lower pressure in the droplet formation chamber will take
place near the exit of the nozzle, where the narrowing of the pipe
system becomes significant.

Using the equations of continuity and momentum conservation, we
can easily find that for laminar flow the mean velocity of the
liquid fluid inside the nozzle ($\bar{v}_\mathrm{noz}$) and the
velocity of jet ($v_\mathrm{jet}$) are related as \cite{Harmon55}
\begin{equation}
v_\mathrm{jet}=\frac{4}{3}\bar{v}_\mathrm{noz}.
\label{eq:vjet_vnoz}
\end{equation}%
The corresponding relationship between the diameter of the fluid inside the
nozzle and the diameter of the jet is given by
\begin{equation}
D_\mathrm{jet}=\frac{\sqrt{3}}{2}D_\mathrm{noz}\approx
0.866D_\mathrm{noz}. \label{eq_d_jet}
\end{equation}
To summarize: for the present nozzle with a 12-$\mu$m inner
diameter we expect a laminar flow inside the nozzle and the jet
size to be $10.4\,\mu$m.

It has been further derived that in turbulent flow case the ratio
between the diameter of the jet and inner diameter of nozzle
($D_\mathrm{jet}/D_\mathrm{noz}$) will be less than $0.866$ if we
solve the momentum-conservation equation with the velocity
distribution of turbulent flow. This could explain why the
experimental measured values of this ratio by other authors are
somewhat smaller (e.g.\ a value of approximately 0.80 was given by
Ref.~\cite{LinSch65}).

The unstable jet will break up into droplets
due to the acoustic excitation. After the satellite droplets merge
into the main droplet, the volume conservation gives
\cite{LinSch65}
\begin{equation}  \label{eq:Dd_Djet}
D_\mathrm{d}=\sqrt[3]{\frac{3}{2}D_\mathrm{jet}^{2}\lambda },
\end{equation}
where $D_\mathrm{d}$ is the droplet diameter, and $\lambda$ is the
wavelength jet turbulence, equivalent to the distance between the
neighboring droplets. From our measured $\lambda=0.368\,$mm we got
$D_\mathrm{d}=39.1\,\mu$m.

\subsubsection{Droplet velocity evolution}
\label{sec:D_vel_ev} When the droplet is detached from the jet, an
extra force will be exerted by the surface tension, $\sigma$,
which pulls the fluid back towards the nozzle. This excess
pressure integrated over the cross section of the jet is
$-P=-2\sigma /D_\mathrm{jet}$. The momentum conservation gives
\cite{Eggers97}
\begin{equation}
v_\mathrm{d}\approx v_\mathrm{jet}-\frac{2\sigma }{\rho
D_\mathrm{jet}v_\mathrm{jet}}, \label{eq:vd_vjet}
\end{equation}
where $v_\mathrm{d}$ is the droplet velocity, and $\sigma$ is the
surface tension of hydrogen liquid.

Now that the droplet velocity was $37.7\,$m/s, we can trace back
to the jet velocity and obtain $v_\mathrm{jet}=37.9\,$m/s.
Recalling the relationship between the jet velocity and the fluid
velocity inside the nozzle, Eq.~(\ref{eq:vjet_vnoz}), numerically,
the fluid velocity at the exit of the nozzle was $28.4\,$m/s, thus
less than $55\,$m/s to fulfil the requirements of laminar flow
(see Sec.~\ref{sec:D_size}).

When the droplet travels in the surrounding of gas mixture, the
gas actually acts as a viscous fluid and exert a frictional drag
force on it. For an isolated spherical droplet, based on Newton's
resistance law, the general expression for the drag force in a gas
surrounding can be written as \cite{Yang00}
\begin{equation}
F_\mathrm{D}=-C_\mathrm{D}\frac{\pi
}{8}\rho_\mathrm{g}D_\mathrm{d}^{2}v_\mathrm{rel}^{2}/C_\mathrm{c}
\label{eq_f_drag}
\end{equation}%
Here $C_\mathrm{D}$ is the drag factor, which has different forms
for different regions of the Reynolds number of the droplet,
$\mathrm{Re}_\mathrm{d}=\rho_\mathrm{g}v_\mathrm{rel}D_\mathrm{d}/
\mu_\mathrm{g}$. Further, $\rho_\mathrm{g}$ is the gas density,
$\mu_\mathrm{g}$ is the gas viscosity, $C_\mathrm{c}$ is
Cunningham non-continuum correction (which is almost 1.0 in our
case because we are working in the continuum region, see
Sec.~\ref{sec:mass_loss}), $A_\mathrm{d}$ is the cross sectional
area of the droplet, i.e.\ $A_\mathrm{d}=\pi D_\mathrm{d}^{2}/4$,
and $v_\mathrm{rel}$ is the relative velocity between the droplet
and the surrounding gas. Since the gas in the droplet formation
chamber can only be pumped away via a $0.8\,$mm narrow vacuum
injection capillary, this limits gas flow such that the resulted
bulk gas velocity in the droplet formation chamber is
insignificantly small. Therefore the background gas in the DFC
could be assumed as stagnant and the relative velocity
$v_\mathrm{rel}$ to equal the droplet velocity $v_\mathrm{d}$. It
should be kept in mind that the surrounding gas is actually a
mixture of hydrogen vapor and helium gas, and the density is the
linear combination of specific volumes, while the viscosity is
deduced according to the summing rule of Wilke \cite{Wilke50}.

For a hydrogen droplet with a velocity of $37.7\,$m/s, the
Reynolds number is $\mathrm{Re}_\mathrm{d}$=34.4. Relevant values
for the background H$_{2}$-He gas at $17\,$K were calculated to
$\rho_\mathrm{g}=0.049\,$kg/m$^{3}$ and $\mu_\mathrm{g}=2.1\times
10^{-6}\,$Pa$\,$s. The hydrogen data for this calculation were
taken from Ref.~\cite{Souers86}. The drag factor equation suitable
for a Reynolds number region of $2<\mathrm{Re}_\mathrm{d}<500$ is
described as \cite{Yang00}
\begin{equation}
C_\mathrm{D}=\frac{24}{\mathrm{Re}_\mathrm{d}}\left[
{1+0.15\mathrm{Re}_\mathrm{d}^{0.687}}\right] .  \label{eq_C_d}
\end{equation}%
This will result in a drag force of about 3000 higher than the gravitational
force exerted on a droplet. Thus the gravity can be omitted and the dynamic
equation for the droplet can then be written as \cite{Yang00}
\begin{equation}
\frac{dv_\mathrm{d}}{dt}=-\frac{18\mu_\mathrm{g}v_\mathrm{d}}{\rho
D_\mathrm{d}^{2}}\left[ {
1+0.15\mathrm{Re}_\mathrm{d}^{0.687}}\right] \label{eq_dv_dt}
\end{equation}
When the droplet passes through the DFC its diameter changes very
little, so as a first order of estimation we can assume that the
droplet diameter is constant. The numerical calculations of
Eq.~(\ref{eq_dv_dt}) show that the relative velocity loss
($1-v_\mathrm{d}/v_\mathrm{d}^{0}$) will be 16\% when the droplet
has travelled an available $7\,$mm distance inside the DFC
(assuming the jet length in Fig.~\ref{fig:DropFormChamber} to be
$1\,$mm). Here $v_\mathrm{d}^{0}$ is the initial droplet velocity
($37.7\,$m/s). The observed relative velocity loss for droplets is
only about $6\%$ from the photos, which is just one-third of the
above value for an isolated droplet. The difference may be
explained by the effect of inter-droplet interactions within the
droplet train, which means that a succeeding droplet in the wake
of a proceeding droplet will endure less drag force
\cite{Devara03}. Further effects of the inter-droplet interaction
will be discussed in the next section.

\subsubsection{Thermodynamic behavior of the droplet} The mass loss
due to evaporation and the connected temperature development of
the droplet can only be achieved by solving the appropriate
thermodynamic equations of the droplet in the DFC. Many
theoretical models exist to describe the heat and mass transfer
processes from micro-droplets \cite{MiHaBe98,ShLeJu00} to the
surrounding for single isolated droplets. However, these models
are of limited benefit in the case of a droplet train where the
inter-droplet separation distance is so small that droplet-droplet
interactions become significant. The intensity of interactions can
be defined in terms of an interaction parameter which is the ratio
between the rate of a transfer process from a droplet in an array
compared to that from a single isolated droplet. Because the  heat
and mass transfer behave very similar, we can reasonably assume
that the interaction parameters for heat and mass transfer rates
are equal, and defined as \cite{Moritz00}
\begin{equation}
\eta =\frac{\Big(\frac{\partial m}{\partial
t}\Big)_\mathrm{arr}}{\Big(\frac{\partial m}{\partial
t}\Big)_\mathrm{iso}},  \label{eq:eta}
\end{equation}
which is the ratio between the mass loss of a droplet in a linear
array (train) relative to that from a single isolated droplet. The
parameter $\eta$ depends only on the dimension-less spacing
between the droplets which is defined as $\epsilon =\lambda /a$,
where $a$ is the droplet radius. From other experiments it has
been reported that the inter-droplet interactions are significant
($\eta \sim 0.3$) for $\epsilon <7.0$ but negligible for $\epsilon
> 20$ \cite{Moritz00,Devara98}. In our case $\epsilon =18.8$, so
we have assumed $\eta $ to be about $0.8$.

\paragraph{Mass loss equation}
\label{sec:mass_loss} Once a droplet is formed, we assume it to
(i) represent all other droplets; (ii) be totally spherical; (iii)
enter a uniform environment with respect to pressure and
temperature; and (iv) be surrounded by its own evaporating gas in a
quasi-steady equilibrium state at the liquid surface so that the
gas pressure follows the saturation line \cite{Souers86}
\begin{equation}
\ln
P_\mathrm{S}=15.52059-\frac{102.7498}{T_\mathrm{s}}+5.338981\times
10^{-2}T_\mathrm{s}-1.105632\times 10^{-4}T_\mathrm{s}^{2},
\label{eq:P_sat}
\end{equation}%
where $P_\mathrm{S}$ is given in Pa if the dimension of the
surface temperature $T_\mathrm{s}$ is K. The heat and mass
transfer process depends on the degree of rarefaction for the
system, which is represented by the dimensionless Knudsen number
defined as $\mathrm{Kn}=l_\mathrm{mfp}/D_\mathrm{d}$ \cite{Yo93}.
Here, $l_\mathrm{mfp}$ is the mean free path, i.e.\ the average
distance between molecule collisions, and $D_\mathrm{d}$ is the
droplet diameter. The ambient gas is a mixture of helium and
hydrogen, so in the simple kinetic gas theory (with the
approximation of rigid spheres) the mean free path is
\cite{Bird02}
\begin{equation}
l_\mathrm{mpf}=\frac{k_\mathrm{B}T_{\infty }^\mathrm{mix}}{\pi \sqrt{2}\big(\frac{1}{2}%
(d_{He}+d_{H_{2}})\big)^{2}P_{\infty }^\mathrm{mix}},
\label{eq_free_path}
\end{equation}
where $k_\mathrm{B}$ is Boltzmann's constant, and
$d_\mathrm{He}=2.15\,${\AA} and $d_\mathrm{H_{2}}=2.71\,${\AA }
are the sphere diameters of helium and hydrogen molecules,
respectively \cite{Lide03}. It is expected that
$l_\mathrm{mfp}\sim 0.04\,\mu$m and $\mathrm{Kn}\sim 0.01$. It is
claimed in Ref.~\cite{Yo93} that at the ``diffusion-controlled
continuum or near-continuum limit'' the evaporation is best
predicted by a diffusion-controlled model if the concentration of
the inert gas is neither too large nor tending to zero when
compared to the evaporated gas. In our case the partial pressures
of helium and hydrogen are of the same order and hence a
diffusion-controlled model should be adequate.

We can thus use Fick's first law of diffusion,
$\mathbf{j}_{\alpha}=- \mathcal{D}_{\alpha \beta }\nabla
\rho_{\alpha }$, to describe the movement of one chemical species
$\alpha $ through a binary mixture of $\alpha$ and $\beta $. Here,
$j_{\alpha }$ is the mass flux rate, the proportionality factor
$\mathcal{D}_{\alpha \beta }$ is the diffusion coefficient, and
$\nabla \rho _{\alpha }$ is the concentration gradient of
$\alpha$. For a spherically symmetric geometry this diffusion
equation can be expressed as \cite{Devara03,ShLeJu00,KoSi01}
\begin{equation}
\frac{\partial m_{\alpha }}{\partial t}=-A\mathcal{D}_{\alpha \beta }\frac{%
\partial \rho _{\alpha }}{\partial r}\Big|_{r=a}\simeq -4\pi a^{2}\mathcal{D}%
_{\alpha \beta }\big(-\frac{\rho _{\infty }-\rho _{a}}{a}\big)=-4\pi a%
\mathcal{D}_{\alpha \beta }(\rho _{a}-\rho _{\infty }).  \label{eq_m_dot}
\end{equation}
where $\frac{\partial m_{\alpha }}{\partial t}$ is the mass
change, $A$ is the droplet surface area $A=4\pi a^2$, $a$ is the
droplet radius ($a=D_d/2$), and $\rho _{a}$ and $\rho _{\infty }$
are the evaporated gas concentrations at $r=a$ and $r=\infty$,
respectively.

For a droplet train moving in the ideal gas mixture,
Eq.~(\ref{eq_m_dot}) can further be expressed as
\begin{equation}
\frac{\partial m_{\alpha }}{\partial t}=-\eta \frac{4\pi a\mathcal{D}%
_{\alpha \beta }}{R_{\alpha }}\big(\frac{P_{a}}{T_{a}}-\frac{P_{\infty }}{%
T_{\infty }}\big),  \label{eq:m_dot_IGL}
\end{equation}
where $R_{\alpha }=R/M_{\alpha}$ is the specific gas constant for
$\alpha$ and, in turn, $\alpha$ and $\beta$ corresponds to
hydrogen and helium, respectively. Here, the assumption of a
quasi-steady equilibrium give $P_{a}=P_\mathrm{S}$, given in
Eq.~(\ref{eq:P_sat}). The diffusion coefficient of gas mixture
$D_{\alpha \beta}$ can be deduced from the kinetic theory as shown
in Ref.~\cite{Bird02}. We have also inserted the earlier discussed
$\eta$-factor to take inter-droplet effects into account.

Since the droplet is non-stationary the mass transfer must be
multiplied by a factor, the Sherwood number, due to the forced
convection. Under the assumption of a negligible internal
circulation inside the droplet, as has been experimentally
investigated in a similar setup and concluded to be reasonable
\cite{Worsnop01}, the Sherwood number for mass transfer is given
by \cite{Renk91}
\begin{equation}
\mathrm{Sh}=2.0+0.552\mathrm{Re}_\mathrm{d}^{1/2}\mathrm{Sc}^{1/3},
\label{eq_sh}
\end{equation}
where the Schmidt number, $\mathrm{Sc}=\mu_\mathrm{g} /\rho
\mathcal{D}_{\alpha \beta }$, is estimated for the gas mixture.

\paragraph{Heat balance equation} \label{sec:HeatBalEq} Inside a
droplet the temperature profile is given by \cite{Devara03}
\begin{equation}
\rho C_\mathrm{P}\frac{\partial T}{\partial
t}=\frac{k}{r^{2}}\frac{\partial }{\partial
r}\Big(r^{2}\frac{\partial T}{\partial r}\Big),  \label{eq:Tr}
\end{equation}%
for $0\leq r\leq a(t)$ and where $\rho $, $C_\mathrm{P}$, and $k$
are the density, the heat capacity and the thermal conductivity of
the droplet, respectively. $T$ is the droplet temperature at the
radial position $r$, at time $t$. Eq.~(\ref{eq:Tr}) applies with
the boundary conditions
\begin{eqnarray}
\frac{\partial }{\partial r}\underbrace{T(0,t)}_{T_\mathrm{c}(t)}
&=&0
\label{eq_bound1} \\
T(a,t) &=&T_\mathrm{s}(t),  \label{eq_bound2}
\end{eqnarray}
where the subscripts $\mathrm{s}$ and $\mathrm{c}$, respectively,
denote surface and center.

At the droplet surface, the heat balance equation can be written
as \cite{Devara03}
\begin{equation}
4 \pi a^{2} k\frac{\partial T}{\partial
r}\Big|_{r=a}=\dot{Q}_\mathrm{vap}+\dot{Q}_\mathrm{conv},
\label{eq:heat_bal}
\end{equation}
with
\begin{equation}
\dot{Q}_\mathrm{vap}=h_\mathrm{vap}\frac{\partial m}{\partial t},
\label{eq_q_vap}
\end{equation}
and
\begin{equation}
\dot{Q}_\mathrm{conv}=\eta 4\pi a^{2}h_{c}(T_{\infty
}-T_\mathrm{s}). \label{eq_q_conv}
\end{equation}
Here, $h_\mathrm{vap}$ is the latent heat of vaporization and
$h_\mathrm{c}=\mathrm{Nu}k/2a$ is the convective heat transfer
coefficient. The Nusselt number $\mathrm{Nu}$ for convective
correction can be expressed as \cite{Renk91}
\begin{equation}
\mathrm{Nu}
=2.0+0.552\mathrm{Re}_\mathrm{d}^{1/2}\mathrm{Pr}^{1/3}
\label{eq_nu}
\end{equation}
where the Prandtl number $\mathrm{Pr}\simeq 0.70$ \cite{Arp98}.

\begin{figure}[htb]
\begin{center}
\includegraphics*[height=10cm]{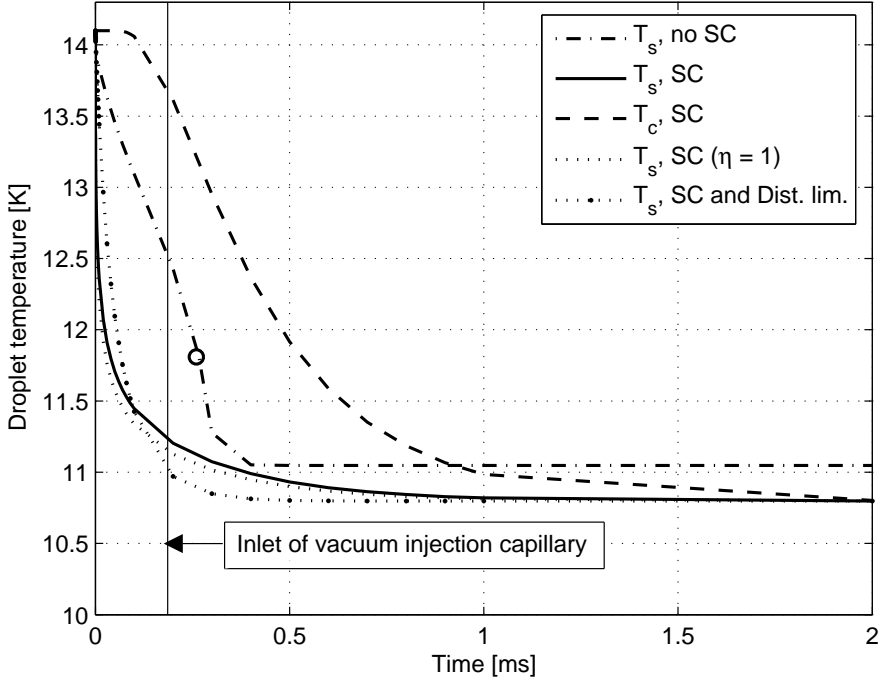}
\end{center}
\caption{Droplet temperature vs.\ time for initial droplets of
$39.1\,\mu$m and $14.1\,$K for different cases. SC is short for
supercooling, whereas $T_\mathrm{s}$ and $T_\mathrm{c}$ denote
surface and center temperatures, respectively. The circle marks
the time where the whole pellet is solid, if no supercooling is
considered. The vertical line at $0.19\,$ms corresponds to the
available time in the droplet formation chamber.}
\label{fig:T_vs_t}
\end{figure}

\paragraph{Droplet temperature results}
With these mass and heat transfer equations, the behavior of the
hydrogen droplet can be obtained using an updated program version
with the numerical iteration method as used earlier in
Ref.~\cite{Tr88}. With an initial droplet diameter of
$39.1\,\mu$m, a velocity of $37.7\,$m/s, and a temperature of
$14.1\,$K, we see in Fig.~\ref{fig:T_vs_t} the droplet temperature
evolution in the DFC for different cases. Our ambient pressure was
$21.3\,$mbar with a partial hydrogen pressure of $8\,$mbar. The
background gas temperature was assumed to be $17\,$K, however, the
variation of this value was shown to be negligible. To simplify
the calculation, the density change which takes place at the
transition between the liquid and solid phase (about $12\%$), was
neglected and instead an average value was used.

When no supercooling occurs, liquid hydrogen transforms to solid
at $T_\mathrm{m}=13.96\,$K and this happens for the surface of the
droplet already after $2.5 \times 10^{-8}\,$s. Assuming a smooth
inward-directed freezing process, the whole micro-sphere would
turn solid at $0.24\,$ms corresponding to a flight path of
$10\,$mm about $3\,$mm into the vacuum injection capillary. This
does not correspond to our experimental observations, because
before a solid pellet is formed the micro-sphere should be
partially frozen and thus fragments of frozen hydrogen should be
visible if the nozzle and thereby the droplet train is tilted to
hit the capillary inlet. Those fragments have never been observed
and therefore we conclude that an additional mechanism must be
present. One distinct possibility is the presence of supercooling
resulting in the corresponding temperature lines in
Fig.~\ref{fig:T_vs_t}. To simulate supercooling for the hydrogen
droplet, we have assumed that all thermodynamic parameters of
liquid hydrogen can be extrapolated to the temperatures lower than
the normal melting temperature. If not stated otherwise, the
droplet train effect is $\eta=0.8$ and the number of layers, to
account for the radial temperature distribution from
Eq.~(\ref{eq:Tr}), is 50. We see that changing $\eta$ to 1.0 (no
effect) make almost no difference. More surprisingly is, maybe,
the small difference between the droplet with constant temperature
distribution (distillation limit model) and the
finite-conductivity model as described above.

\paragraph{Nucleation rate of hydrogen} Generally, the reason for supercooling
is that the crystallization has no place to take root and the
molecular motion prevents the substance from freezing. The
temperature $T_\mathrm{n}$ at which a droplet of volume
$V_\mathrm{d}$ will start to freeze, is determined by the relation
\cite{Hi78,Be00}
\begin{equation}  \label{eq_Nn}
N_\mathrm{n}=-V_\mathrm{d}\int_{T_\mathrm{n}}^{T_\mathrm{m}}
\frac{\Gamma(T)}{\dot{T}}\, dT\simeq 1,
\end{equation}
where $N_\mathrm{n}$ is the number of nuclei and $\dot{T}$ is the
temperature change. The time dependence from the latter makes it
possible to find a corresponding static formula,
\begin{equation}
\tau \simeq \frac{1}{\Gamma V_\mathrm{d}} ,  \label{eq_gama}
\end{equation}
where $\tau$ is the lifetime of the supercooled droplet before
freezing. Eq.~(\ref{eq_gama}) has been experimentally investigated
for parahydrogen (p-H$_{2}$) and, to some extent, also normal
hydrogen (n-H$_{2}$) \cite{Se86,Ma87}. They measured $\tau$ and
$V_\mathrm{d}$ to obtain the nucleation rate $\Gamma$, in an
helium environment of $\sim 15\,$bar. Our case a comparable low
ambient pressure, $\sim 20\,$mbar, and we will therefore neglect
this effect such that the theory of classical nucleation rate can
be used. Furthermore, we note that the nucleation temperature of
interest is not too far away from the melting temperature, so by
using the notation of Ref.~\cite{Ma83}, we get
\begin{equation}
\Gamma
=\frac{n_\mathrm{L}k_\mathrm{B}T}{h}e^{-(\phi_\mathrm{LS}+\delta
F_{0})/k_\mathrm{B}T}, \label{eq_nucl_rate}
\end{equation}
with
\begin{equation}
\delta F_{0}=\frac{16\pi }{3}\frac{\alpha_\mathrm{LS}^{3}}{
n_\mathrm{S}^{2}(f_\mathrm{L}-f_\mathrm{S})^{2}}.
\label{eq_free_eng}
\end{equation}%
and
\begin{equation}
f_\mathrm{L}-f_\mathrm{S}\approx
0.99(T_\mathrm{m}-T)-0.029(T_\mathrm{m}-T)^{2}, \label{eq_delt_f}
\end{equation}
for the case of parahydrogen. Here $k_\mathrm{B}$ is Boltzmann's
constant, $h$ is Planck's constant, $n_\mathrm{L}$ is the number
of molecules per unit volume in the liquid phase,
$\phi_\mathrm{LS}=\phi_\mathrm{L}=45k_\mathrm{B}$ is the
activation energy for self-diffusion in the the liquid, $\delta
F_{0}$ is the maximum free energy required to form a small solid
sphere in the liquid, $\alpha_\mathrm{LS}$ is the liquid-solid
surface free energy which was determined to be $0.874\times
10^{-3}\,$J/m$^{2}$ (according to the best fit of experimental
data \cite{Se86}), $n_\mathrm{S}$ is the number of molecules per
unit volume in the solid phase, and $f_\mathrm{L}$ and
$f_\mathrm{S}$ are, respectively, the free energies per molecule
in the liquid and solid phases.

To turn back to normal hydrogen we use the results of
Ref.~\cite{Ma87} where it is pointed out that the nucleation rate
for n-H$_{2}$ was measured to be $\sim 10^{3}$ times higher than
that of p-H$_2$, possibly due to the ``onset of rotational
ordering in the solid phase''. From this and
Eqs.~(\ref{eq_nucl_rate}), (\ref{eq_free_eng}), and
(\ref{eq_delt_f}) the approximate nucleation rates for n-H$_{2}$
can be obtained, see Tab.\ \ref{tab:Nucl}.

\begin{table}[htb]
\centering
\begin{tabular}{cccc}
\hline\hline
$T$ $[$K$]$ & $\Gamma$ $[$cm$^{-3}$s$^{-1}$$]$ & $T$ $[$K$]$ & $\Gamma$ $[$cm%
$^{-3}$s$^{-1}$$]$ \\ \hline
$9.8$ & $4.9 \times 10^{11}$ & $10.8$ & $1.7 \times 10^{0}$ \\
$10.0$ & $1.3 \times 10^{10}$ & $11.0$ & $2.4 \times 10^{-4}$ \\
$10.2$ & $1.7 \times 10^{8}$ & $11.2$ & $4.3 \times 10^{-9}$ \\
$10.4$ & $1.1 \times 10^{6}$ & $11.4$ & $5.0 \times 10^{-15}$ \\
$10.6$ & $2.5 \times 10^{3}$ & $11.6$ & $1.4 \times 10^{-22}$ \\ \hline\hline
\end{tabular}%
\caption{Assumed nucleation rate for hydrogen (n-H$_{2}$) at low
(zero) ambient pressure, based on Eq.~(\protect\ref{eq_nucl_rate})
and the experimental factor $\sim 10^{3}$ \protect\cite{Ma87}.}
\label{tab:Nucl}
\end{table}

When nucleation occurs the latent heat of fusion is liberated, and
the droplet warms up to the so-called recalescence arrest
temperature $T_\mathrm{r}$ \cite{Ma87,MaApLa89}. Eventually the
heat stored in the droplet is exhausted, and it rapidly freezes to
solid (pellet).

For a $\sim39\,\mu$m-diameter droplet the volume is $3 \times
10^{-8}\,$cm$^{3}$. According to Eq.~(\ref{eq_gama}) and values
from Tab.~\ref{tab:Nucl}, the needed droplet temperature is
$9.8\,$K for the nucleation to start inside the DFC at about
$0.19\,$ms. As seen in Fig.~\ref{fig:T_vs_t} this temperature will
never be reached, not even if the DFC was extended to several
decimeters. In fact, the lowest droplet temperature of $10.8\,$K
corresponds to a nucleation rate which is 11 orders of magnitude
to low for a nucleation trigger. Given that the nucleation rate
may have a large uncertainty, this huge order anyway points to
that freezing in the DFC is impossible because the droplet is not
``supercooled enough'' to experience a phase transition to solid.
Besides, if we consider the supercooling-case for the droplet
center and surface temperature, respectively, i.e.\ the dashed and
solid lines in Fig.~\ref{fig:T_vs_t}, at the correspondent time
for the inlet of the vacuum injection capillary, we actually have
$T_\mathrm{c}=13.6\,$K and $T_\mathrm{s}=11.2\,$K. That is greater
than $10.8\,$K and thus an even lower nucleation rate.

In the calculations we have changed all related working
parameters, such as the vapor pressure, the background gas
pressure, and the background gas temperature. It is found that
only when the pressures of vapor and background gas have been
decreased to as low as $3\,$mbar and $10\,$mbar, respectively, the
droplet can finally reach $9.8\,$K in the DFC. This will however
be very hard to achieve experimentally since the hydrogen vapor
pressure results from the evaporation mechanism itself and the
pumping power in the DFC is limited by the narrow vacuum injection
capillary.

To trigger the nucleation and thereby the freezing other means are
needed. A controlled introduction of impurities might be the
solution, e.g.\ as mentioned in Ref.~\cite{Se86} by heating a fine
tungsten filament such that impurities at the surface are boiled
off to imply a heterogenous crystallization. Or, maybe supersonic
excitation can be used to decrease the degree of supercooling even
further \cite{Inada01}, and thereby increase the nucleation rate.

\subsection{Pellet evaporation}
Consider the possibility that the micro-spheres are still in
liquid phase after they have passed the vacuum injection
capillary. Using liquid parameters in the equations to come, it
turns out that the droplet will reach a temperature well below
$9.8\,$K within $0.5\,$ms, almost independent of chosen diameter
or initial temperature. From the discussions on nucleation rate we
thus conclude that if the droplet does not transform to a pellet
inside the vacuum injection capillary, it will do that within some
centimeters afterwards. Therefore all equations in vacuum are
subscripted $\mathrm{p}$ for pellets.

In vacuum the pellet will evaporate in the molecular flow regime
($\mathrm{Kn}\gg 1$), and the evaporation rate is given by the
(classical) Hertz--Knudsen formula. However, a revised version
also takes the bulk velocity of the vapor in vacuum into account
\cite{YtOs96}, such that
\begin{equation}
\frac{\partial m_{\alpha }}{\partial t}=-1.668 A\frac{
P_\mathrm{S}-P_{\infty }}{\sqrt{2\pi R_{\alpha }T}},
\label{eq:m_dot_corrected}
\end{equation}%
where the solid-vapor saturation pressure follows \cite{Souers86}
\begin{equation}
\ln P_\mathrm{S}=9.2458-\frac{92.610}{T_\mathrm{s}}+2.3794\ln
T_\mathrm{s}. \label{eq:P_sat_sol}
\end{equation}
The main heat source in the vacuum chamber, i.e.\ after the
injection capillary, is the thermal radiation from the walls of
the vacuum pipe. The pellets are exposed to the wall surface with
a temperature of about $298\,$K. On the basis of the
Stefan--Boltzmann law combined with Kirchhoff's law, the thermal
radiation can be written as
\begin{equation}
\dot{Q}_\mathrm{rad}=4\pi a^{2}\varepsilon \sigma
(T_\mathrm{w}^{4}-T_\mathrm{s}^{4}) \label{eq:q_dot_rad}
\end{equation}
where $\varepsilon $ is the emissivity of the pellet (assumed to be 0.5), $%
\sigma $ is the Stefan--Bolzmann constant ($5.67\times
10^{-8}\,$Wm$^{-2}$K$^{-4}$), $T_\mathrm{w}$ is the pipe-wall
temperature ($298\,$K), and $T_\mathrm{s}$ is the surface
temperature of the pellet.

At the pellet surface, the energy balance gives
\begin{equation}
4\pi a^{2}k_\mathrm{p}\frac{\partial T_\mathrm{p}}{\partial
r}\mathstrut
\Big|_{r=a}=\dot{m}h_\mathrm{sub}+\dot{Q}_\mathrm{rad}
\label{eq:pel-serf}
\end{equation}
where $k_\mathrm{p}$ is the thermal conductivity of the pellet and
$h_\mathrm{sub}$ is the latent heat of sublimation. The energy
balance equation inside the pellet is the same with that in the
droplet formation chamber, i.e.\ Eq.~(\ref{eq:Tr}) with the
boundary conditions Eqs.~(\ref{eq_bound1}) and (\ref{eq_bound2}),
but for which all involved parameters should be interchanged to
those of solid hydrogen.

Once the pellet is formed, we can assume that the initial
temperature of the pellet will be equal to the triple point
temperature $13.96\,$K when they are just consolidated from
supercooled droplets. Of course other values are in principle
possible, but due to the rapid temperature decrease the chosen
temperature is really not of importance. The mass loss and
temperature evolution can be calculated by solving the mass
transfer and heat balance, i.e.\ Eqs.~(\ref{eq:Tr}),
(\ref{eq:m_dot_corrected}), and (\ref{eq:pel-serf}). The results
are shown below in Figs.~\ref{fig:PelletTempChange} and
\ref{fig:PelletMassLoss}. It turns out that pellets will cool down
to about $6\,$K within $10\,$cm of travel in the vacuum chamber.
As seen the lines overlap for the different cases. The relative
mass loss during this process is about 14\%. After that, i.e.\ on
the way down to the skimmer and further to the interaction point,
the pellet mass loss rate will be maintained at a rather low
constant value. This is very important for a pellet target
operation inside the vacuum of a storage ring which should not be
spoiled.

\begin{figure}[htb]
\begin{center}
\includegraphics*[height=7cm]{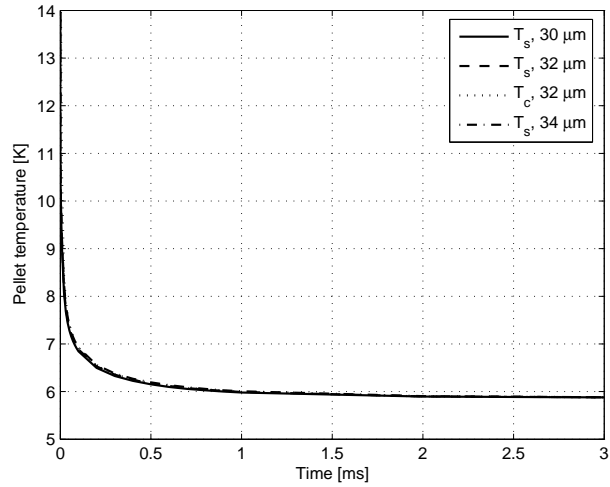}
\end{center}
\caption{The pellet temperature decreases quickly and reaches a
final temperature of $5.9\,$K. Since there is no obvious
difference between the four curves, we conclude that the
temperature change is more or less independent on the initial
diameter and the radial position. $T_\mathrm{s}$ and
$T_\mathrm{c}$ denote surface and center temperatures,
respectively.} \label{fig:PelletTempChange}
\end{figure}

\begin{figure}[htb]
\begin{center}
\includegraphics*[height=7cm]{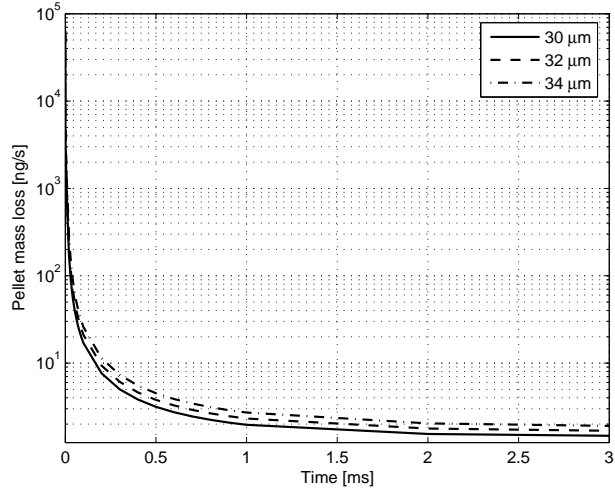}
\end{center}
\caption{The pellet mass loss becomes almost constant after about 1 ms and
varies only slightly due to the initial pellet diameter. By far, most of the
mass is lost during the first millisecond.}
\label{fig:PelletMassLoss}
\end{figure}

\begin{table}[htb]
\centering
\begin{tabular}{cccc}
\hline\hline
Initial diameter & $\frac{dm}{dt}(t=0\,$ms$)$ & $\frac{dm}{dt}(t=1\,$ms$)$ &
$\frac{dm}{dt}(t=5\,$ms$)$ \\
$[\mu$m$]$ & $[$ng/s$]$ & $[$ng/s$]$ & $[$ng/s$]$ \\ \hline
$30$ & $-5.7 \times 10^{4}$ & $-2.0$ & $-1.5$ \\
$32$ & $-6.4 \times 10^{4}$ & $-2.3$ & $-1.7$ \\
$34$ & $-7.3 \times 10^{4}$ & $-2.7$ & $-1.9$ \\ \hline\hline
&  &  &  \\
&  &  &
\end{tabular}%
\caption{Mass change in vacuum for different pellet sizes and times.}
\label{tab:Mass}
\end{table}

\section{Summary}
We have described how uniform spaced and sized droplets are formed
in the droplet formation chamber by an acoustical excitation
method. From experimental observations we know that the
micro-spheres are in liquid phase at the inlet of the vacuum
injection capillary and completely frozen at the skimmer. Thus
they must freeze in between. In the absence of supercooling the
calculations show that they would be only partially frozen at the
capillary inlet, in contradiction to the observations. However,
with supercooling present the experimental observations can be
explained. Our calculations show that the nucleation rate is 11
orders of magnitude too low to trigger a pellet freezing at all.
It is desirable to achieve frozen pellets within the DFC, and
therefore the introduction of impurities or the usage of
supersonic excitation should be considered. The expected advantage
of pellets formed at an earlier stage is an improvement of the
survival ratio, and thus better working conditions for a pellet
target.

The total relative mass loss from the formation of droplets to
pellets in the reaction chamber is estimated to be 30\%,
corresponding to a $\sim10$\% decrease in diameter. The pellet
size itself is actually further decreased due to the density
change for hydrogen from liquid to solid. As for an original
droplet with a diameter of $39\,\mu$m, the final pellet size is
expected to be about $33\,\mu$m.

With the current setup, the droplets freeze to pellets either in
the vacuum injection capillary or a few centimeters afterwards. In
either case, the pellet equilibrium temperature of about $6\,$K is
reached within $10\,$cm after the capillary. Our calculations show
that the mass loss has converged to an almost constant value at
this point. The mass loss results in an unwanted gas load. Most of
the gas load will come from the mass lost during those $10\,$cm as
well as from the skimmed off pellets eventually breaking up and
evaporating in the vacuum chamber above. However, this gas load
can be pumped away by appropriate vacuum pumps before and after
the skimmer. Thus the gas load to consider for the vacuum system
of the storage ring in which the target is installed, really
originates from the converged value of the mass loss. For our
pellet target this mass loss is between $1.5$ and $1.9\,$ng/s per
pellet.


\ack We are very grateful to Curt~Ekstr{\"o}m and Inti~Lehmann,
whose comments considerably improved the quality of the paper. One
of us ({\"O}.N.) acknowledges the financial support from GSI,
Darmstadt, Germany, and Z.-K.L.\ was supported by the Chinese
Academy of Sciences.





\end{document}